\documentclass[letterpaper,twocolumn,10pt]{article}
\usepackage{usenix2019,epsfig,endnotes}
\usepackage{cleveref}
\usepackage[table,xcdraw]{xcolor}
\usepackage{tikz}
\usepackage{subfig}

\newcommand*\circled[1]{\tikz[baseline=(char.base)]{
		\node[shape=circle,draw,inner sep=0.5pt] (char) {#1};}}

\begin{document}
	
	\date{}
	
	\title{\Large \bf A Survey on Tiering and Caching in High-Performance Storage Systems}
	
	\author{
		{\rm Morteza Hoseinzadeh}\\
		University of California, San Diego
	} 
	
	\maketitle
	
	\thispagestyle{empty}

	\subsection*{Abstract}
		
        Although every individual invented storage technology made a big step towards perfection, none of them is spotless. Different data store essentials such as performance, availability, and recovery requirements have not met together in a single economically affordable medium, yet. One of the most influential factors is price. So, there has always been a trade-off between having a desired set of storage choices and the costs. To address this issue, a network of various types of storing media is used to deliver the high performance of expensive devices such as solid state drives and non-volatile memories, along with the high capacity of inexpensive ones like hard disk drives. In software, caching and tiering are long-established concepts for handling file operations and moving data automatically within such a storage network and manage data backup in low-cost media. Intelligently moving data around different devices based on the needs is the key insight for this matter. In this survey, we discuss some recent pieces of research that have been done to improve high-performance storage systems with caching and tiering techniques.
	
	\section{Introduction\protect\footnote{Parts of this section is taken from my published papers~\cite{yang2017autotiering,hoseinzadeh2014reducing,shengan2019ziggurat}.}}
\begin{figure}[t]
			\centering
			\includegraphics[width=3in]{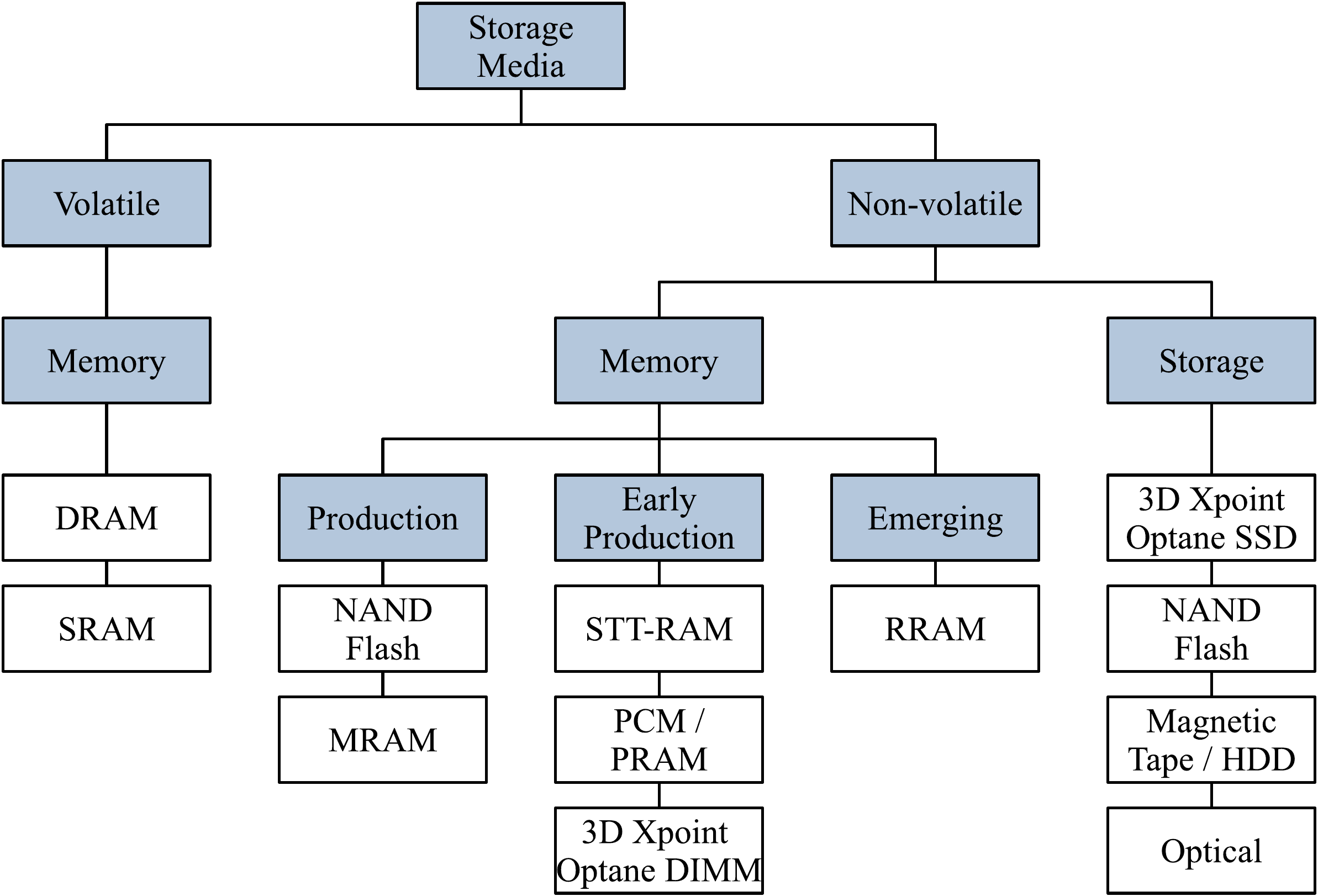}
			\caption{Memory technologies}
			\label{fig:types}
		\end{figure}

		With the advancement in the computing and networking technologies especially around the Internet, and emerging tremendous number of new data sources such as Internet of Things (IoT) endpoints, wearable devices, mobile platforms, smart vehicles, etc., enterprise data-intensive analytics input is now scaled up to petabytes and it is predicted to be exceeding 44 zettabytes by 2020~\cite{turner2014digital}. Concerning this rapid data expansion, hardware has been endeavoring to provide more capacity with higher density supporting high-performance storage systems. Figure~\ref{fig:types} represents available and emerging storage technologies as of today. In terms of storage technology, Hard Disk Drives (HDD) is now supplanted by fast, reliable Solid State Drives (SSD). Additionally, one-time emerging persistent memory devices are now going to be available in the market as Intel launched Optane DIMMs~\cite{optanedimm}. Price-wise, when new technologies become available, dated technologies become cheaper. Nowadays, SSDs are very common such that they are being used as All-Flash Arrays (AFA) in data centers~\cite{yang2017autotiering}. However, storage IO is still the biggest bottleneck on large scale data centers. As shown in~\cite{andersen2010rethinking}, the time consumed to wait for I/Os is the primary cause of idling and wasting CPU resources, since lots of popular cloud applications are I/O intensive, such as video streaming, file sync, backup, data iteration for machine learning, etc.

	\begin{table*}[th] 
		\centering
		\caption{Comparison of different storage technologies~\cite{arulraj2015let,volos2011mnemosyne,wikibon2017,dulloor2014system}}
		\label{tbl:str}
		\small
		\begin{tabular}{|r|c|c|c|c|c|c|}
			\hline 			& STT-RAM & DRAM & NVDIMM & Optane SSD$^\dagger$ & NAND SSD$^\ddagger$ & HDD \\ \hline
			Capacity* 		& 100s of MBs & Up to 128GB & 100s of GBs & Up to 1TB & Up to 4TB & Up to 14TB \\ \hline
			Read Lat. 		& $6ns$ & $10-20ns$ & $50ns$ & $9\mu s$ & $35\mu s$ & $10ms$ \\ \hline
			Write Lat. 		& $13ns$ & $10-20ns$ & $150ns$ & $30\mu s$ & $68\mu s$ & $10ms$ \\ \hline
			Price 			& \$1-3K/GB & \$7.6/GB & \$3-13/GB & \$1.30/GB & \$0.38/GB & \$0.03/GB \\ \hline
			Addressability 	& Byte & Byte & Byte/Block & Block & Block & Block \\ \hline
			Volatility 		& Non-Volatile & Volatile & Non-Volatile & Non-Volatile & Non-Volatile & Non-Volatile \\ \hline
			\multicolumn{7}{l}{\tiny $^\dagger$Intel Optane SSD 905P Series (960GB) (AIC PCIe x 4 3D~XPoint) ~~~~ $^\ddagger$Samsung 960 Pro 1TB M.2 SSD with 48-layer 3D NAND (Source: Wikibon) ~~~~ *Per module} 
		\end{tabular}
	\end{table*}
		
		To solve the problem caused by I/O bottlenecks, parallel I/O to multiple HDDs in a Redundant Array of Independent Disks (RAID) becomes a common approach.
		However, the performance improvement from RAID is still limited. Therefore, lots of big data applications strive to store intermediate data to memory as much as possible such as Apache Spark.
		Unfortunately, memory is too expensive, and its capacity is minimal (e.g., 64$\sim$128GB per server), so it alone is not able to support super-scale cloud computing use cases.
		Some researches propose making use of NVM-based SSDs like 3D~XPoint Optane DIMM~\cite{3dxpoint,optane} and PCM-based DIMMs~\cite{hoseinzadeh2014reducing,hoseinzadeh2016spcm,optane} instead of DRAM to provide high density and non-volatility. But, these storage devices are not matured enough to be instantly used as the main memory and are still very expensive.
		
		Caching and Tiering have been used for a long time to hide long latency of slow devices in the storage hierarchy. In the past, high-end HDDs such as 15k RPM where used as the performance tier and low-end HDDs such as 7200 RPM served as the capacity tier~\cite{muppalaneni2000multi}. Today, NAND-Flash SSDs replaces fast HDDs, and while low-end HDDs are obsolete, high-end HDDs are used for capacity requirements. Soon, modern storage technologies such as NVM will break with the past and change the storage semantics. As in device level, today's high-speed SSDs are equipped with a write buffer as in Apple's Fusion Drive~\cite{fusion}. In system level, almost all file systems come with a page cache which buffers data pages in DRAM, and letting applications have access to the contents of the files. Using persistent memory as a storage medium, some file systems skip the page cache~\cite{xu2016nova}.
		In application level, lots of big data applications strive to store intermediate data to memory as much as possible such as Apache Spark. However, NVM is not economically affordable to be used as a large enterprise storage system, and SSDs suffer from limited write endurance.
		
		In this survey, we discuss several studies on caching and tiering solutions for high-performance storage systems. In section~\ref{sec:bkgnd}, we give a short background of storage devices and their technologies. Section~\ref{sec:caching} will investigate several research studies on caching solutions followed by section~\ref{sec:tiering} which discusses several papers on storage tiering solutions. At the end of section~\ref{sec:tiering}, we briefly introduce Ziggurat, which is developed in our group. Finally, section~\ref{sec:conc} concludes the paper.

	\section{Background}
		\label{sec:bkgnd}
        This section briefly covers background information of individual technology parts in the computer memory hierarchy. We also discuss the counterpart pieces of hardware and software required for networking them together.
		
		\subsection{Memory Hierarchy}
		
        Based on the response time, the memory hierarchy is designed to separate the computer storage into an organized multi-level structure aiming to enhance the overall performance and storage management. Different types of storage media are designated as levels according to their performance, capacity, and controlling technology. In general, the lower level in the hierarchy, the smaller its bandwidth and the larger its storage capacity. There are four primary levels in the hierarchy as follows~\cite{toy1986book}.
		
		\subsubsection{Internal}
            On-chip memory cells such as processor registers and caches fall into this level. To provide the highest performance, architects use storage technologies with the lowest response type such as SRAM, Flip-Flops, or Latch buffers. Embedded DRAM is another technology which is used in some application specific integrated circuits (ASIC)~\cite{hamzaoglu2014edram}. In recent years, some emerging technologies such as spin-torque transfer random access memory (STT-RAM) has received attention for the last level cache~\cite{sun2011multi,smullen2011relaxing}. They not only provide low response time, but they also offer high density and persistence.

			Notice that there are multiple sub-levels in this level of the memory hierarchy. Processor register file, which has the lowest possible latency, resides in the nearest sub-level to the processor followed by multiple levels of caches (i.e., L1, L2, and so on). Although in symmetric multi-processor (SMP) architecture caches may be private or shared amongst the cores, they are still considered on the same level in the hierarchy. 
			
		\subsubsection{Main}
            The primary storage or the main memory of the computer system temporarily maintains all code and data (partial) of the running applications including the operating system. At this level of the hierarchy, the capacity is more important compared with the internal levels. The whole code and data of running applications settle at this level. Although the storage capacity in this level is much larger than the internal level, the performance should also be high enough to enable fast data transfer between the main and internal levels. Using the spacial and temporal locality, the memory controller manages to move bulks of data back and forth between the last level cache and the main memory via the address and data bus. In contrast with internal levels in which data can be accessed in bytes, unit access of data is a cache or memory line (usually 64 Bytes).
            
            DRAM technology has been long used as the best candidate for this level. Other technologies such as phase change memory (PCM)~\cite{lee2009architecting,lee2010phase,hoseinzadeh2014spcm} have been introduced as a scalable DRAM alternative with the ability to persist data. 3D~XPoint~\cite{3dxpoint} has been successfully prototyped and announced. Detailed information on the storage technologies can be found in~\cref{sec:tech:mem}.
			
		\subsubsection{Secondary Storage}
			The secondary storage or the on-line mass storage level is composed of persistent block devices to store massive data permanently. In contrast with the two levels above, the storage is not directly accessible by the processor. Instead, the storage media are connected to the processor via IO ports. Solid State Drives (SSD), Hard Disk Drives (HDD), and rotating optical devices are examples of secondary storage media. When a process is being executed, the processor submits an IO request to the block device via an IO BUS such as PCIe, IDE, or SATA in order to load a chunk of data (usually a block of 4KB) into a specific location in the main memory using the Direct Memory Access (DMA) feature.
		
		\subsubsection{Tertiary Storage}
            The tertiary storage or off-line bulk storage includes any kinds of removable storage devices. If accessing the data is under control of the processing unit, it is called tertiary storage or near-line storage. For example, a robotic mechanism mounts and dismounts removable devices on demand. Otherwise, it is called off-line storage, when a user physically attaches and detach the storage media. In some storage classifications, tertiary storage and off-line storage are distinguished. However, we consider them identical in this paper. The rest of this section will discuss the most related technologies and their characteristics.

	\subsection{Technology}
	\label{sec:tech}
        The main factor that makes storage media different from each other is their technologies. Throughout the computer history, memory technologies have been evolved vastly. Figure~\ref{fig:types} represents currently available and emerging technologies at a glance. Generally, the computer memory system can be classified into volatile and non-volatile memories. Traditionally, non-volatile memories which usually fall in secondary and tertiary storage groups, are used to store data permanently. In contrast, volatile memories are usually used as caches to temporarily maintain close to the processor because of their high performance. Nevertheless, their usage may switch often. For example, a high-end SSD may be used as a cache for slow storage devices. Likewise, recently emerged storage class memories can be used as a non-volatile media to permanently store data in spite of being in the primary storage place. Table~\ref{tbl:str} compares different computer storage technologies.

		\subsubsection{Memory Technology}
		\label{sec:tech:mem}
		
            SRAM and DRAM have been long known as the primary technologies served as the processor's internal cache and the system's main memory, respectively. Due to the nature of an SRAM cell, it can retain information in no time. An SRAM cell is composed of two back-to-back inverters. In its standby state, these two inverters keep reinforcing each other as long as they are supplied. One of them represents bit data, and the other one corresponds to the inverted value of the bit data. While reading, a sense amplifier reads the output ports of the inverters and find which one has a higher voltage and determines the stored value. Although SRAM is almost as fast as a logic gate circuit, its density is too low as its electronic structure is made of at least four transistors. Additionally, it is CMOS compatible, so, integrating SRAM cells in the processor's die is possible. On the other hand, a DRAM cell comprises only one transistor and a capacitor. In contrast with SRAM which statically keeps data, DRAM requires refreshing the data due to the charge leakage nature of the capacitor. The density of DRAM is much higher than SRAM, but it is not CMOS compatible. So, integrating DRAM in the processor's die is not easy. Also, it requires larger peripheral circuitry for read and write operations. Since reading from a DRAM cell is destructive, a write should happen following each read to restore the data. Overall, the higher capacity with a lower cost of DRAM made it the best candidate for the primary memory, so far. However, DRAM has faced a scaling wall because it uses electric charge in capacitors to maintain data. So, while technology scaling, not only the reliability of a capacitor dramatically drops, but also there would be cell-to-cell interference. Not to mention that the active power consumption of refresh overhead is another challenging issue.
			 
			Many emerging technologies have been investigated to address the scaling issue among others. Researchers have been seeking a reliable solution for a byte-addressable and power efficient alternative to DRAM. Spin-Transfer Torque RAM (STT-RAM) is one of the high-performance solutions~\cite{hosomi2005novel}. Having a fixed layer and a free layer of ferromagnetic material, it stores bits in the form of high and low resistance property of the fixed layer based on the spin orientation of the free layer. Although it provides higher performance comparing with DRAM along with non-volatility which voids refreshing, its expensive costs make it an unfordable option of DRAM replacement. Its super high density and low power consumption make it a potential candidate for on-CPU cache technology.
			
			Nonetheless, Phase Change Memory (PCM) is another emerging technology which is more promising than the others. It stores digital information in the form of resistance levels of a phase change material which ranges from little resistance of its crystalline state to very high resistance of its amorphous state~\cite{lee2009architecting}. As shown in table~\ref{tbl:str}, PCM has a lower performance compared with DRAM, especially in write operations. It also can endure a smaller number of writes and requires refreshing to prevent resistance drift. There is a body of research focusing on addressing these issues~\cite{hoseinzadeh2014spcm,yoon2014efficient,nair2015reducing}.
			
			Notwithstanding, PCM is one of the best options to be used as a storage class memory technology, and solid-state drives. Table~\ref{tbl:str} shows the beneficiary of PCM and 3D~XPoint devices over NVMe driver. Connecting to the memory bus, they provide near DRAM performance while having a large capacity of a storage device. This type of memory technology is recognized as Storage Class Memory (SCM) which can be categorized as memory type (M-SCM, Persistent Memory, or NVM) with fast access latency and low capacity (as in 3D~XPoint DIMM), or storage-type (S-SCM) with high capacity and low access latency (as in Optane SSD, see section~\ref{sec:tech:store})~\cite{yamada2016optimal}.
		
		\subsubsection{Storage Technology}
		\label{sec:tech:store}
			Besides the internal and the main memory, permanent data should reside in some storage device to be accessed on demand. For a long time, Hard Disk Drives (HDD) have been playing this role. An HDD consists of rigid rapidly rotating disks held around a spindle and a head which relocates using an actuator arm. Digital data is stored in the form of transitions in magnetization of a thin film of magnetic material on each disk. The electromechanical aspect of HDD and the serialization of the stored data make HDD orders of magnitude slower than the mentioned non-volatile memory technologies. However, its low price and extremely high density make it a good candidate for secondary and tertiary storage levels. According to table~\ref{tbl:str}, the capacity of an HDD can be 1000x larger than DRAM while the operational latency is roughly $10^6$ times slower.
			
			Solid State Drives (SSD) offer higher performance, shock resistance, and compact storage at the cost of higher prices by using Flash technology. A Flash cell consists of a MOSFET with one word-line control gate and another floating gate. It keeps data in the form of electrical switch in the floating gate which can be programmed to be on or off. Whether the networking of the MOSFETs resembles a NAND or a NOR logic, it is called NAND-Flash or NOR-Flash SSD. The read operation is as simple as reading the bit-line while charging the word-line. However, writing to a flash cell requires erasing a large portion (MBs) of storage area using tunnel release and put data afterward with tunnel injection. SSDs may use traditional protocols and file systems such as SATA, SAS, NTFS, FAT32, etc. There are also some interfaces such as mSATA, m.2, u.2, and PCIe, and some protocols such as NVMe that are specifically designed for SSDs. The capacity of NAND-flash based SSD ranges from 128GB to 100TB, and the performance can be up to 10GB/s. Despite all benefits that a NAND-Flash SSD provides, its lifespan is limited to $10^4$ writes per cell. Intel and Micron recently shipped Optane SSD with the new technology of 3D~XPoint~\cite{3dxpoint} that offers longer lifespan and higher performance. A 3D~XPoint cell preserves data based on the change of bulk resistance~\cite{clarke2015bulk}. Due to the stackable cross-gridded arrangement, the density of 3D~XPoint is much higher than traditional non-volatile memory technologies. Intel also announced 3D~XPoint DIMM form-factor which can provide memory band-with for non-volatile storage.

		\subsubsection{Mass Storage Dilemma}
\begin{figure}[t]
				\centering
				\includegraphics[width=3in]{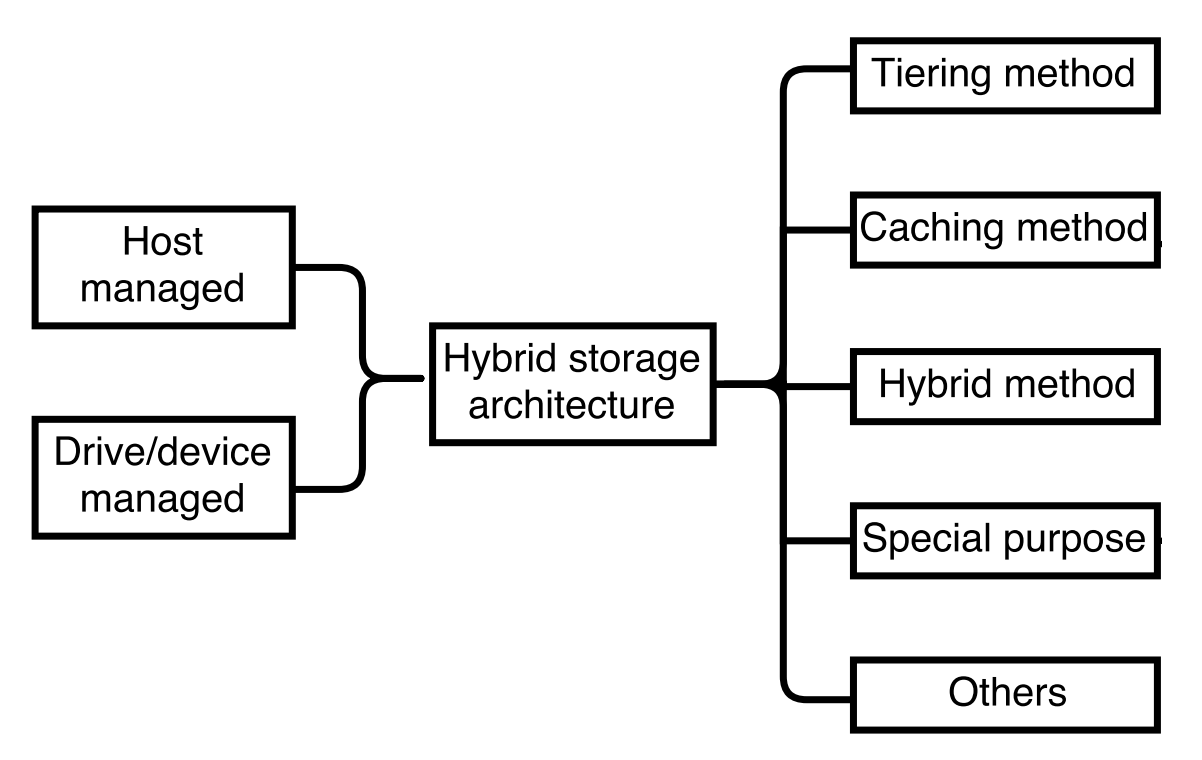}
				\caption{Hybrid storage architectures~\cite{niu2018hybrid}}
				\label{fig:hyb}
			\end{figure}

            The technologies mentioned above are engaged in different levels on the memory hierarchy. In one hand, organization of the storage system in the hierarchy can vary based on data intensity. In the other hand, the pace of data growth in data centers and cloud-storage service providers mandates server administrators to seek a high-performance mass storage system which requires a data management software running on top of networked storage devices and server machines. Therefore, choosing one technology to design a massive storage system is not the best solution. So, data center experts opt to develop a hybrid storage system~\cite{niu2018hybrid}. Figure~\ref{fig:hyb} depicts the overall categories of the hybrid storage architectures. In this study, we focus on a host-managed tiering and caching methods.
	
            The exponentially expanding digital information requires fast, reliable, and massive data centers to not only archive data, but also rapidly process them. So, high-performance and large capacity are both required. However, the portion of digital information with different values may not be even. The IDC report~\cite{turner2014digital} predicts that with the speed of doubling every two years, the size of the digital universe might exceed 44 zettabytes ($2^{70}$) by 2020. This tremendously extensive information is not being touched equally. While the cloud will touch only 24\% of the digital universe by 2020, 13\% will be stored in the cloud, and 63\% may not be touched at all~\cite{turner2014digital}. The speed of data that requires protection, which is more than 40\%, is even faster than the digital universe itself. So, the major of data usually resides in cheaper, more reliable, and larger devices and the minor of it which is still not processed is preserved in fast storage media. Therefore, a hybrid storage system with a caching/tiering mechanism will be undoubtedly required.
	
	\section{Storage Caching Solutions} 
	\label{sec:caching}
        With the aim of alleviating the long latency of slow devices, a caching mechanism can be used in a hybrid storage system. There are two main principles in caching subsystems: 1) while keeping the original data in the moderate levels of the hierarchy, a \textit{copy} of under-processing data resides in the cache; and 2) the lifetime of data in the cache layer is short, and it is meant to be \textit{temporary}. The performance of the storage system with caching is chiefly influenced by four factors~\cite{niu2018hybrid}:

		\begin{enumerate}
			\item Data allocation policy essentially controls the data flow and determines the usefulness of the cache, accordingly. The distribution of the data among multiple devices is reflected by the caching policy, such as read-only, write-back, etc.
			\item The translation, depending on its mechanism, may also influence the performance. In a hybrid storage system, the same data may be kept in different locations in multiple devices, and each copy of the data should be addressable. The address translation mechanism is important to be fast for data retrieval, and compact for metadata space usage.
			\item An accurate data hotness identification method is necessary for better cache utilization. It helps to prevent cache pollution with unnecessary data, and consequently, improving the overall performance by instantly providing hot data.
			\item The cache usage efficiency is another important factor which is influenced by the scheduling algorithm for managing the queues, synchronization, and execution sequence.
		\end{enumerate}
		
		A caching mechanism can be managed by either in hardware by the device or in software by the host operating system (see figure~\ref{fig:hyb}). Device-managed caching systems are beyond the scope of this study, so we focus on host-managed methods. With a host-manage caching mechanism, the host may use separate devices to enhance the performance. One of the most common cases is using SSD as a cache because of its high performance as opposed to slow HDDs, and high capacity compared with DRAM. Besides SSDs, emerging Non-Volatile Memory (NVM) devices are promised to be involved in storage caching mechanisms. In this section of the paper, we discuss a few storage caching techniques including using either SSD or SCM as a storage cache.

		\begin{figure}[t]
			\centering
			\includegraphics[width=3in]{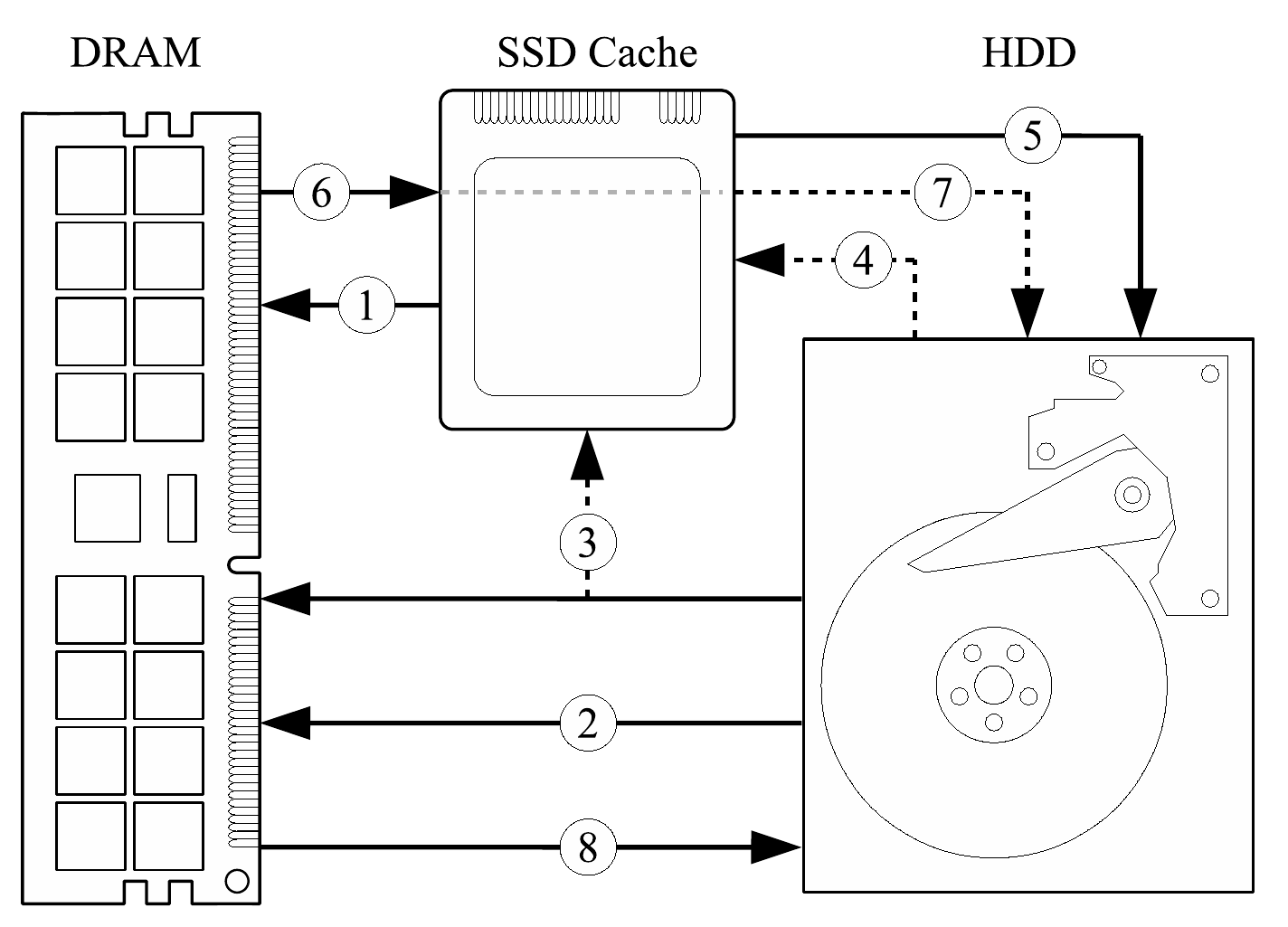}
			\caption{Dataflows in SSD caches}
			\label{fig:ssd-cache}
		\end{figure}
		
		\subsection{SSD as a Cache}
		Covering the performance gap between the disk drive and the main memory, SSD devices have been widely used for caching slow drives. Figure~\ref{fig:ssd-cache} shows common data flows in a caching system using an SSD device. \circled{1} happens when the read request completes within the SSD cache without involving the HDD. If the requested block is not in the SSD, the HDD may be accessed to retrieve data in DRAM via \circled{2}, and if it is identified as a hot data, it is going to be cached in SSD via \circled{3}. A background process which executes a hot data identification may migrate data from the HDD to the SSD via \circled{4} regardless of not being requested. A flush command or a write-back can copy dirty blocks back to the HDD in \circled{5}. A write operation may be completed directly in SSD when the block is already there as in \circled{6}, and whether the cache uses a \textit{write-through} or \textit{write-back} policy, the dirty block can be copied in HDD via \circled{7}. The \textit{write-through} policy in SSD caches is obsolete as it is designed for volatile caches in which dirty blocks should be persisted at some point. In case of using the \textit{read-only} or \textit{write-around} policy, all new write operations are performed by \circled{8} directly in HDD. Based on the caching policy, data may flow through these paths.
		
		\subsubsection{SSD as a Read-Only Cache} 
		\label{sec:ro}
		Upon arrival of a new write request in a read-only cache architecture~\cite{liu2013molar,tai2015sla,dai2015etd,zhao2016towards,yang2013hec,meng2014vcacheshare} where the accessing block is not located in SSD, the request is completed by successfully recording it to HDD via \circled{8}. When it was already cached in SSD for priority read operations, the request is considered as completed only after updating the HDD copy of data and discarding the SSD copy, successfully. This kind of cache architecture helps the durability of the SSD device as the writing traffic to the SSD is limited to fetching data from HDD. Meanwhile, the cache space can be better utilized for reading operations, and it might improve the overall read performance which, unlike write operations, is on the critical path. However, the SSD lifespan is still vulnerable to the cache updating policy. If the data selection is not accurate enough, the cache might be polluted with unnecessary data, and a Garbage Collection (GC) process or a replacement mechanism should run to make space for demanding data. This process may incur a write overhead to the SSD and reduce the lifespan. 
		
		The replacement algorithm is essential to alleviate the writing pressure on SSD cache. Section~\ref{sec:repl} will discuss more on common algorithms. Besides, the block hotness identification also affects the SSD lifespan vastly. MOLAR~\cite{liu2013molar} determines the data hotness based on the control metric of demotion count, and place the evicted blocks wisely from the tier-1 cache (DRAM) to the tier-2 cache (SSD). Using the I/O patterns of applications on an HPC system, \cite{zhao2016towards} proposes a heuristic file-placement algorithm to improve the cache performance. Since the applications in an HPC is more mechanized as opposed to end-user applications which have an unpredictable I/O pattern, assuming foreknown patterns is not far from being realistic. To understand the I/O pattern, a distributed caching middleware detects and manipulates the frequently-accessed blocks in the user level. 
		
		\subsubsection{SSD as a Read-Write Cache}
		Due to its non-volatility feature, SSD caches do not use a write-through policy to keep the original data up-to-date, in contrast with DRAM caches which are volatile and need to be synchronized or persistent. So, an SSD R/W cache may only employ a write-back or flushing mechanism. Using SSD as an R/W cache to improve the performance in terms of both read and write operations is very common~\cite{huang2016improving,lee2015effective,liu2010raf}. In such architectures, new writes are performed in the SSD cache as shown in figure~\ref{fig:ssd-cache}:\circled{6}, and they will be written back to the disk later. Since there are two versions of the same data, a periodic flush operation usually runs to prevent data synchronization problem. Although using an R/W SSD cache normally improves the storage performance, when the cache is nearly full, it fires the GC process to clean invalid data which may interfere the main process and degrade the performance. Meanwhile, if the workload is write-intensive with a small ratio of data reuse, the HDD may be under a heavy write load which prevents the disk to have long idle periods. This fact wards off the SSD flushing process and impose extra performance overhead to the system. However, SSD can keep data permanently, thus flushing all write data is not necessary. So, a write-back cache policy can improve storage performance. Nevertheless, an occasional flush operation at the cost of small performance degradation is required in case of SSD failure problem. Furthermore, the SSD limited write endurance is another issue which is more problematic in R/W caches comparing with read-only caches. Notice that the random write in an SSD device is roughly tenfold slower than the sequential write and causes excessive internal fragmentation. Many algorithms~\cite{chen2017duplication,huang2016improving,dai2015etd,liang2016elastic} and architectures~\cite{yang2013hec,liu2010raf,oh2012caching} have been design to alleviate the write traffic and control the GC process in SSD caches. 
		
		Random Access First (RAF)~\cite{liu2010raf} cache management extends the SSD lifespan by splitting the SSD to read and write caches. The former one maintains random-access data evicted from file cache with the aim of reducing flash wear and write hits. The latter one is a circular write-through log to respond to write requests faster and perform the garbage collection. A monitoring module in the kernel intercepts page-level operations and sends them to a dispatcher who is a user-level daemon performing random-access data detection and distributes the operations among the caches.
		In \cite{oh2012caching}, balancing the read and write traffics in two different parts of the cache is beneficial for both performance and SSD lifespan. These parts can use different technologies such as DRAM, NVM, or SSD. In section~\ref{sec:ro} we described SSD as an RO cache in which the write traffic may go to the DRAM cache. In other designs, SSD may be used as a write cache for HDD.
		
		\subsubsection{SSD Caches in Virtualization Environments}
		In a virtualization environment with multiple Virtual Machines (VM) running with different IO patterns, the randomness of write operations is a pain-point for SSD flashes. To reduce the number of random writes, \cite{lee2015effective} proposes a cache scheme in which they adopt the idea of log-structured file systems to the virtual disk layer and convert the random writes to sequential writes. Leveraging Sequential Virtual Disks (SVD) in a virtual environment of a home cloud server with multiple virtual machines (VM) in which synchronous random writes dominate, it uses SSD drives completely sequentially to prolong its lifespan while improving the performance. 
		vCacheShare~\cite{meng2014vcacheshare} is an SSD cache architecture on a virtual cluster which simply skips SSD cache for write operations. By tracing the IO traffic of each virtual disk and analyzing them periodically, vCacheShare optimally partitions the SSD cache for each of the virtual disks.
		
		\subsubsection{Deduplication in SSD Caches}
		
		For expanding SSD's lifetime, deduplication is one of the most effective ways. Some research studies~\cite{dai2015etd,chen2017duplication} prevent writing data to the SSD drive if the contents were already cached. For instance, \cite{chen2017duplication} reduces the number of writes to the SSD by avoiding duplicated data in a virtualization environment in which the high integration of VMs can introduce a lot of data duplication. Using a hash function (SHA-1), data signature will be calculated upon a data fetch after a cache miss, and if the signature was already in the cache, the address would be mapped to the content, and it saves one write operation.
		
		CacheDedup~\cite{li2016cachededup} is an in-line deduplication mechanism for Flash caching in both client machines and servers. This design is complementary to Nitro~\cite{li2014nitro} with a set of modifications on the architecture and the algorithms for deduplication-aware cache management. The benefits of deduplication are not only a better utilization of the cache space but also it helps to increase the hit-ratio. Additionally, since the flash devices are limited in write endurance, it also delays wearing out the device by avoiding excessive writes due to duplicate data.

		\begin{figure}[h]
			\centering
			\includegraphics[width=3in]{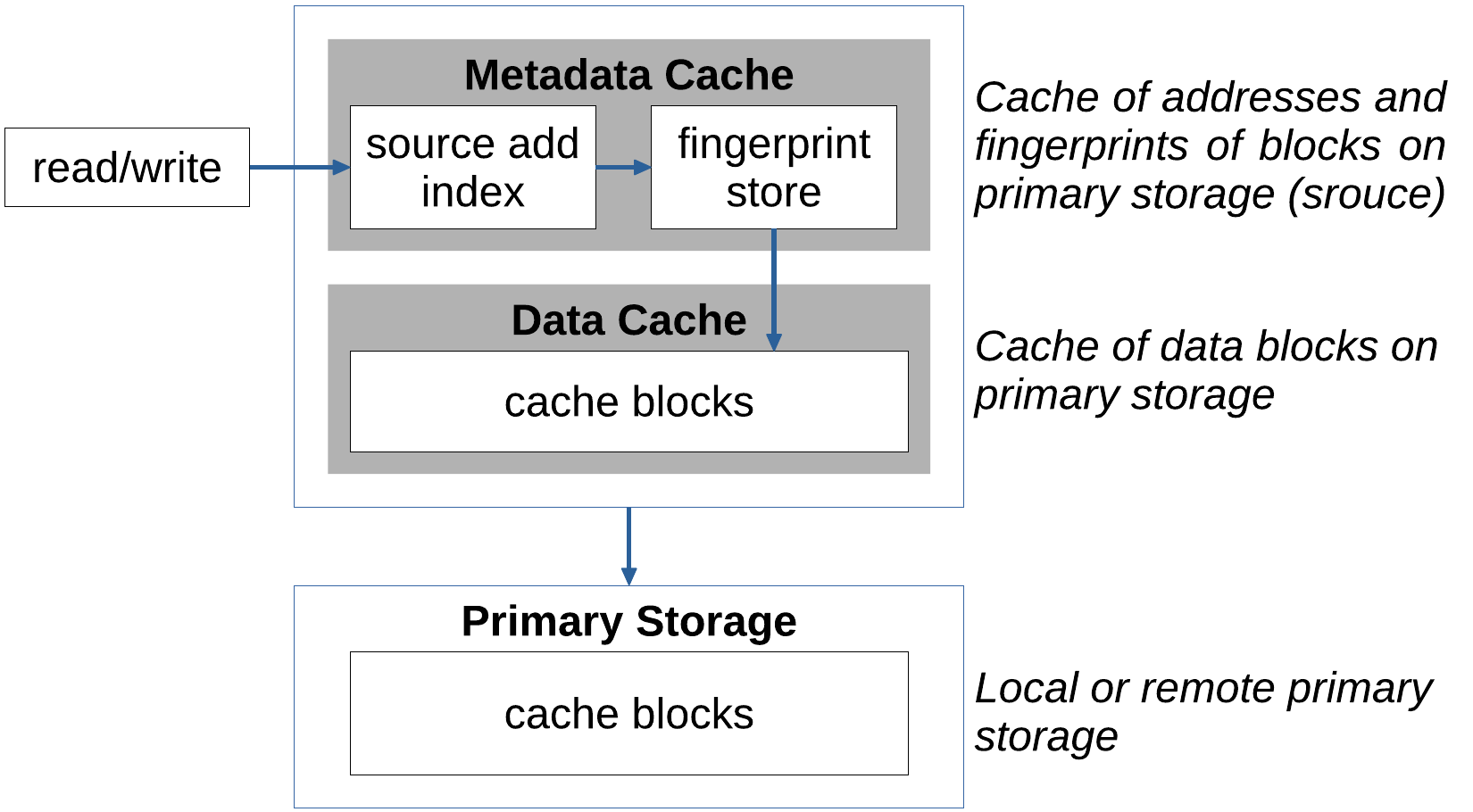}
			\caption{Architecture of CacheDedup~\cite{li2016cachededup}}
			\label{fig:cachededup}
		\end{figure}
		
		As shown inf figure~\ref{fig:cachededup}, CacheDedup is composed of two data structures: Metadata Cache and Data Cache. The Metadata Cache maintains the information for tracking the footprints of source addresses in the primary storage. It has a source address index table and a footprint store. When a read/write operation comes up, the corresponding footprint index is obtained from the source-to-index mapping, and then the footprint-to-block-address mapping gives the block address of corresponding contents in the Data Cache. Since the source-to-cache address space has a many-to-one relationship due to eliminating duplicate data, the size of mappings is not bounded to the size of the cache. Also, to prevent re-fetching data from the primary store, CacheDedup keeps the historical fingerprints for those blocks that have already been evicted.
		CacheDedup can be deployed on both client and server machines. When it is running on a client machine, it can better hide the Network I/O for duplicate data and hence get better performance for applications. In server side, multiple clients may request for the same data, and CacheDedup can help data reduction. Notice that in the server side there should be cache coherence protocol over the network to maintain data consistency. Although the proposed design is described all in software, the authors claim that it can be embedded in the Flash Translation Layer in the hardware device, as well. The described system works with block I/O level referring source block addresses, but it also can be used in file system level with (file handler, offset) tuple.
		One of the main parts of the design is the replacement algorithm. There are two algorithms: D-LRU and D-ARC. The details can be found in section~\ref{sec:repl}. D-ARC algorithm is more complicated than D-LRU. D-ARC has a scan-resistant nature which prevents single-accessed data to pollute the cache capacity. Although both algorithms can be used in CacheDedup, D-ARC achieves better performance while D-LRU is simple. Both algorithms have the no-cache-wastage property, i.e., it doesn’t allow orphaned address and orphaned data blocks at the same time. This study shows the improvement on cache hit ratio, I/O latency, and the number of writes sent to the cache device.

		\subsubsection{SSD as a Cache for SMR Drives}
		Although random write to SSD is slower than sequential write, yet it is an order of magnitude faster than random writes in a Shingled Magnetic Recording (SMR) device such as HDD. Therefore, to benefit from the high-capacity and low \$/GB of SMRs and the high performance of SSDs, a hybrid storage system may redirect all random writes to the SSD cache and leave the sequential writes to the SMR, as in~\cite{xiao2016hs,wang2017larger,lu2016design}.
		
		\begin{figure}[h]
			\centering
			\includegraphics[width=3in]{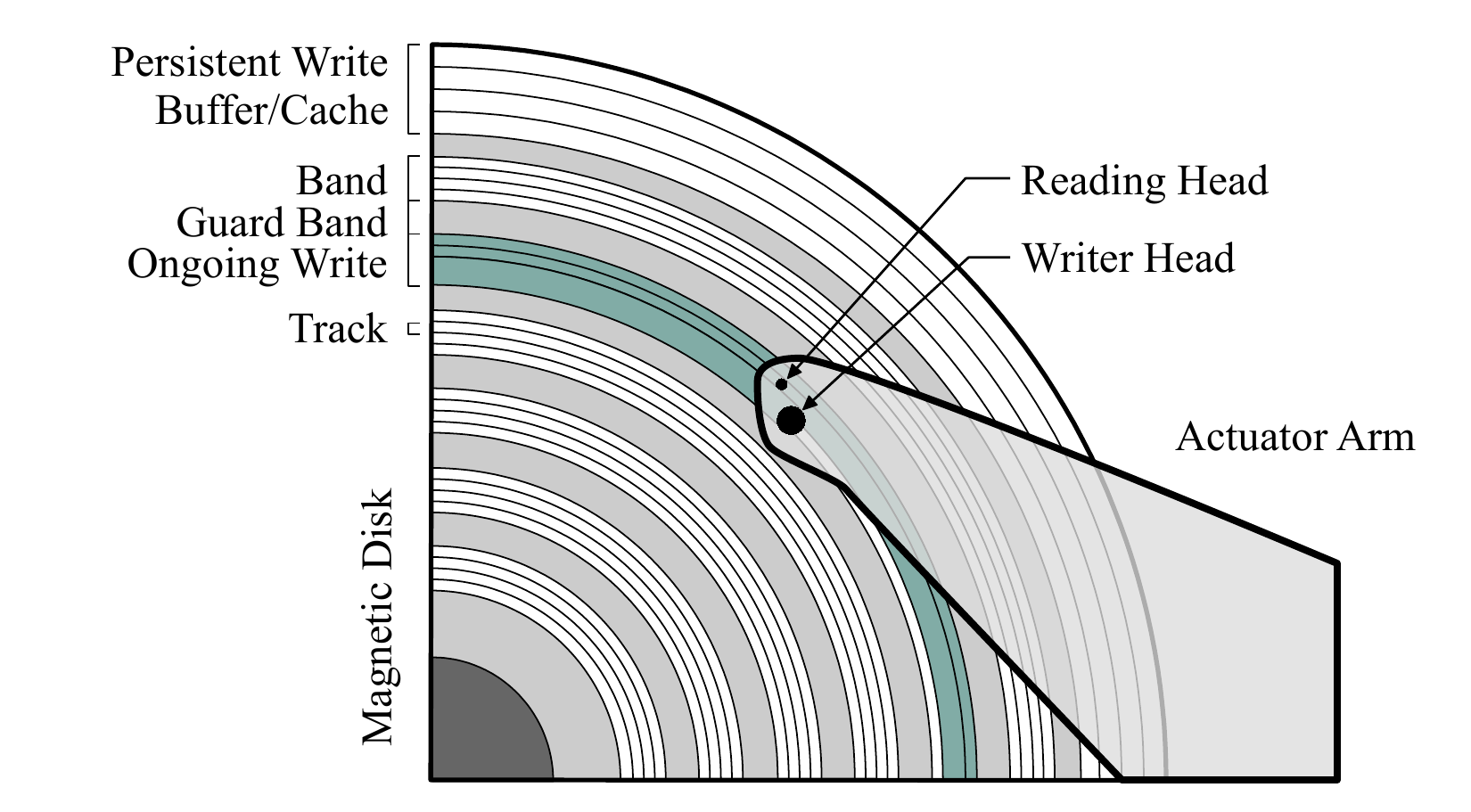}
			\caption{A magnetic disk in a SMR device under operation of writing new data on a whole band.}
			\label{fig:shingle}
		\end{figure}
		
		The writing mechanism is depicted in figure~\ref{fig:shingle}. In an SMR drive, writing to a magnetic \textit{track} partially overlaps a previously written track and makes it narrower to provide higher density. This is because of the physical limitations of the writing head which is not as delicate as the reading head and remarks a wider trail while writing onto a magnetic disk. As imagined, a random write destroys adjacent tracks, and they all should be rewritten by a read-modify-write (RMW) operation. In the SMR architecture, a \textit{band} is a bunch of consecutive tracks grouped and separated from adjacent bands by a narrow \textit{guard band}. Random write requires an RMW operation on a whole band. So, a persistent cache which is either a flash buffer or some non-overlapped track on the magnetic disk is used to buffer writes before writing them to the corresponding band. HS-BAS~\cite{xiao2016hs} is a hybrid storage system based on band-awareness of SMR disk to improve the performance of Shingled Write Disk (SWD or SMR disk) with sustained random writes by taking SSD as a writing cache for SWD. To make use of SWD devices in a RAID system, \cite{lu2016design} proposes three architectural models of using SSDs as caches for SWDs. With this option, a RAID system may provide more storage capacity at lower cost with the same performance or even slightly better. Partially Open Region for Eviction (PORE)~\cite{wang2017larger} caching policy is another use of SSD as a cache for SMR devices. It considers the SMR write amplification due to the Logical Block Address (LBA) wide range in addition to the popularity for replacement decision making. To put it simply, SSD handles random writes and flushes sequentially to the SMR device.
		
		\subsection{NVM Storage Cache}
		The advent of NVM technologies, as described in section~\ref{sec:tech:mem}, allow persistent data operations at near-DRAM latencies, which is an order of magnitude faster than SSD. A study~\cite{lee2018empirical} on using NVM as an I/O cache for SSD or HDD reveals that the current I/O caching solution cannot fully benefit from the low-latency and high-throughput of NVM. Recent researches have been trying to overcome the complexity of using NVM as a Direct Access (DAX) storage device and using it as a cache for SSD/HDD~\cite{bhaskaran2013bankshot,fan2014h,kim2014evaluating,wei2017transactional}.
		In recent years, Intel provided a Persistent Memory Development Kit (PMDK)~\cite{pmdk} which provides several APIs to access the persistent memory from the user level directly.
		NVM Bankshot~\cite{bhaskaran2013bankshot} is a user-level library exposing the NVM by implementing caching functions to the applications and bypassing the kernel to lower the hit latency. However, PMDK outperforms Bankshot in many ways as it is more recent. 
		Most NVM technologies can endure orders of magnitude more writes comparing with NAND SSD, but still limited. They also provide an in-place byte-size update which is way faster than RMW operations in SSDs. With these features, most of DRAM caching policies can be used as NVM-based cache with significant modifications for carefully managing the write traffic. 
		
		Hierarchical ARC (H-ARC)~\cite{fan2014h} cache is an NVM-based cache that optimizes ARC algorithm to take four states of recency, frequency, dirty, and clean into account and split the cache first into the dirty-/clean-page caches and then split each part into recency-/frequency-page cache. Based on a similar mechanism as ARC (see section~\ref{sec:repl}), it adapts the sizes of each section, hierarchically in each level. So, H-ARC keeps dirty pages with higher frequency longer in the cache.
		I/O-Cache~\cite{fan2015cache} also uses NVM as a buffer cache for HDDs which coalesces multiple dirty blocks into a single sequential write. This technique is also used in many other NVM-based designs~\cite{kwon2017strata,shengan2019ziggurat}. 
		Transactional NVM disk Cache (Tinca)~\cite{wei2017transactional} aims to achieve crash consistency through transactional supports while avoiding double writes by exploiting an NVM-based disk cache. Leveraging the byte addressability feature of NVM, Tinca maintains fine-grained cache metadata to enable copy-on-write (COW) while writing a data block. Tinca also uses a role switch method in which each block has a role and can be either a \textit{log block} in ongoing committing transactions, or a \textit{buffer block} in a completed transaction. With the two of COW and role switch mechanisms, Tinca supports a commit protocol to coalesce and write multiple blocks in a single transaction.
		
		\subsection{Cache Replacement Algorithms}
		\label{sec:repl}
		To keep the most popular blocks in the cache, several general-purpose and domain-specific algorithms have been designed. In general, the majority of these algorithms are based on two empirical assumptions that are temporal locality and skewed popularity~\cite{huang2016improving}. The former assumes that the recently used blocks are most likely going to be requested shortly again. The latter supposes that some blocks are more frequently accessed comparing with the others. Accordingly, the well-known mechanism of Least-Recently-Used (LRU) and Least-Frequently-Used (LFU) have been created and commonly used for data replacement in caches because of their simplicity and $O(1)$ overhead. Unlike CPUs, the storage applications may not be interested in the temporal locality since there is a page cache in the DRAM which adequately manages the locality. Also, a simple search operation over the entire storage space may flush all popular blocks in the cache and replace them with seldom accessed ones. There are many more advanced algorithms have been proposed to address this issue which is mostly general-purpose.
		
		\subsubsection{General Purpose Algorithms}
		The Frequency-Based Replacement (FBR)~\cite{robinson1990fbr} algorithm benefits from both LRU and LFU algorithms. It keeps LRU ordering and decides primarily upon the frequency count of the blocks in a section. Its complexity ranges from $O(1)$ to $O(log_2n)$ according to the section size. Using the aggregation of recency information for block referencing behavior recognition, Early Eviction LRU (EELRU)~\cite{smaragdakis1999eelru} aims to provide an on-line adaptive replacement method for all reference patterns. It would perform LRU unless many recently fetched blocks had just been evicted. In that case, a \textit{fallback} algorithm either evicts the LRU block or the $e^{th}$ MRU one, where $e$ is a pre-determined recency position. The Low Inter-reference Recency Set (LIRS)~\cite{jiang2002lirs} algorithm takes \textit{reuse distance} as a metric for dynamically ranking accessed blocks. It divides the cache into a Low Inter-reference Recency (LIR) for most highly ranked blocks and a High Inter-reference Recency (HIR) for other blocks. When an HIR block is accessed, it goes to the LIR, and when LIR is full, the lowest ranked block from LIR turns into the highest ranked HIR block. With the aim of removing cold blocks quickly, 2Q~\cite{sasha19942q} uses one FIFO queue $A1_{in}$ and two LRU lists of $A1_{out}$ and $A_m$. A first accessed block comes into $A1_{in}$, and upon eviction, it goes to $A1_{out}$. Reusing the block promotes it to $A_{m}$. Similarly, Multi-Queue~\cite{zhou2001mq} algorithm uses multiple LRU queues of $Q_0, ..., Q_{m-1}$ where the block lifetime in $Q_j$ is longer than $Q_i$ ($i<j$) as a block in $Q_i$ is hit at least $2^i$ times. Adaptive Replacement Cache (ARC)~\cite{megiddo2003arc} divides the cache space into $T_1$ and $T_2$, where $T_1$ stores one-time accessed blocks whereas $T_2$ keeps the rest of the blocks. Two ghost caches $B_1$ and $B_2$ maintains the identifiers of evicted blocks from $T_1$ and $T_2$, respectively, whereas t. Using $B_1$ and $B_2$, the sizes of $T_1$ and $T_2$ is dynamically adjusted by a dividing point $P$ to balance between recency and frequency which is tuned according to hit rates. 
		
		\subsubsection{Domain Specific Algorithms}
		Base on the write performance of SSD and its lifetime issue, SSD caches usually consider two factors: 1) keeping dirty pages longer in the cache to avoid fetching a page more than once, and 2) avoiding cache space pollution with low popular blocks.
		Clean First LRU (CFLRU)~\cite{park2006cflru} splits the cache space into a \textit{clean-page} cache and a \textit{dirty-page} cache, and evicts only from the clean-page cache unless there is no clean page left. This basic algorithm tries to keep dirty pages longer in the cache, but yet it ignores skipping one-time access pages.
		Lazy ARC (LARC)~\cite{huang2016improving} is designed explicitly for SSD caches to prevent write overheads and prolong the SSD lifespan. It filters the seldom accessed blocks and skips caching them. Similar to 2Q and ARC, it considers the fact that blocks which are hit recently at least twice are more likely to be popular. It has a ghost cache to keep the identifiers of the first accessed blocks. If a block from the ghost cache is reaccessed, it is considered popular and placed in the cache. Since it prevents unnecessary writes to the SSD, it can be also categorized as a data hotness identification method. 
		The Second-level ARC (L2ARC)~\cite{gregg2008zfs,leventhal2008flash} is also optimized for SSD caches as it reduces the number of writes to the device. It has been used in the Solaris ZFS file system. It uses SSD as the second level cache of the in-DRAM ARC cache to periodically fill it with the most popular data contents of the DRAM cache.
		With a large space overhead, SieveStore~\cite{pritchett2010sievestore} keeps the information of the miss count of every block in the storage system, and only allows those blocks with large miss count to be cached in SSD. Similar algorithms are used in some enterprise products such as Intel Turbo Memory~\cite{matthews2008intel}.
		
		Similar to ARC, Duplication-aware ARC (D-ARC)~\cite{chen2017duplication,li2016cachededup} consists of four LRU caches. D-ARC uses cache block contents or fingerprint instead of addresses. Based on the high or low levels of dirty ratio and the reference count of blocks, it partitions data in four groups and always evicts least referenced and dirtiest cache blocks. Hence, the removed block is more likely the most unpopular one which is not going to be reused in the near future, and the SSD would not eject it up to the point that it is no longer required. This will reduce the write bandwidth to the SSD device and save extra writes due to false evictions.
		To the same end, Expiration-Time Driven (ETD)~\cite{dai2015etd} cache algorithm delays a cache block eviction to its expiration time, and instead of updating the cache on a miss, it the evicts a block when it is expired, and then chooses a replacement form a list of candidate blocks.
		D-LRU~\cite{li2016cachededup} is a duplication-aware LRU which consists of two separate LRU policies. First, it inserts the address $x$ in Metadata Cache using LRU. Second, the corresponding fingerprint of address $x$ is inserted into Data Cache using the other LRU.
		
		PORE~\cite{wang2017larger} is another domain-specific policy which is beneficial in SSD-SMR hybrid storage systems. It splits the SMR LBA range into \textit{Open} and \textit{Forbidden} regions. The replacement policy may only evict dirty blocks in the open region. The written back blocks are stored in the SMR write buffer or persistent cache for subsequent writing to the corresponding band. The open region is periodically changed to cover all dirty blocks across the SMR LBA range. This algorithm helps to avoid writing on random bands which significantly destroys the performance.
		
		\section{Storage Tiering Solutions}
		\label{sec:tiering}
		In the past, high-end and low-end HDDs were used as the performance and the capacity tiers, respectively. Nowadays, many types of storage media with different characteristics and capacities are used in a multi-tiered storage system. The main difference between caching and tiering is that in a caching system, a copy of data is kept in the cache while in a tiering system, the original data migrates between multiple tiers via two operations of promotion and demotion. Data is classified based on the application needs and characteristics of available tiers, usually into hot and cold. The hot data resides in the performance tier leaving the cold data to stay in the capacity tier. Considering multiple factors such as randomness, transfer speed, etc., there might be more than two tiers.
		
		\begin{figure}[h]
			\centering
			\includegraphics[width=3in]{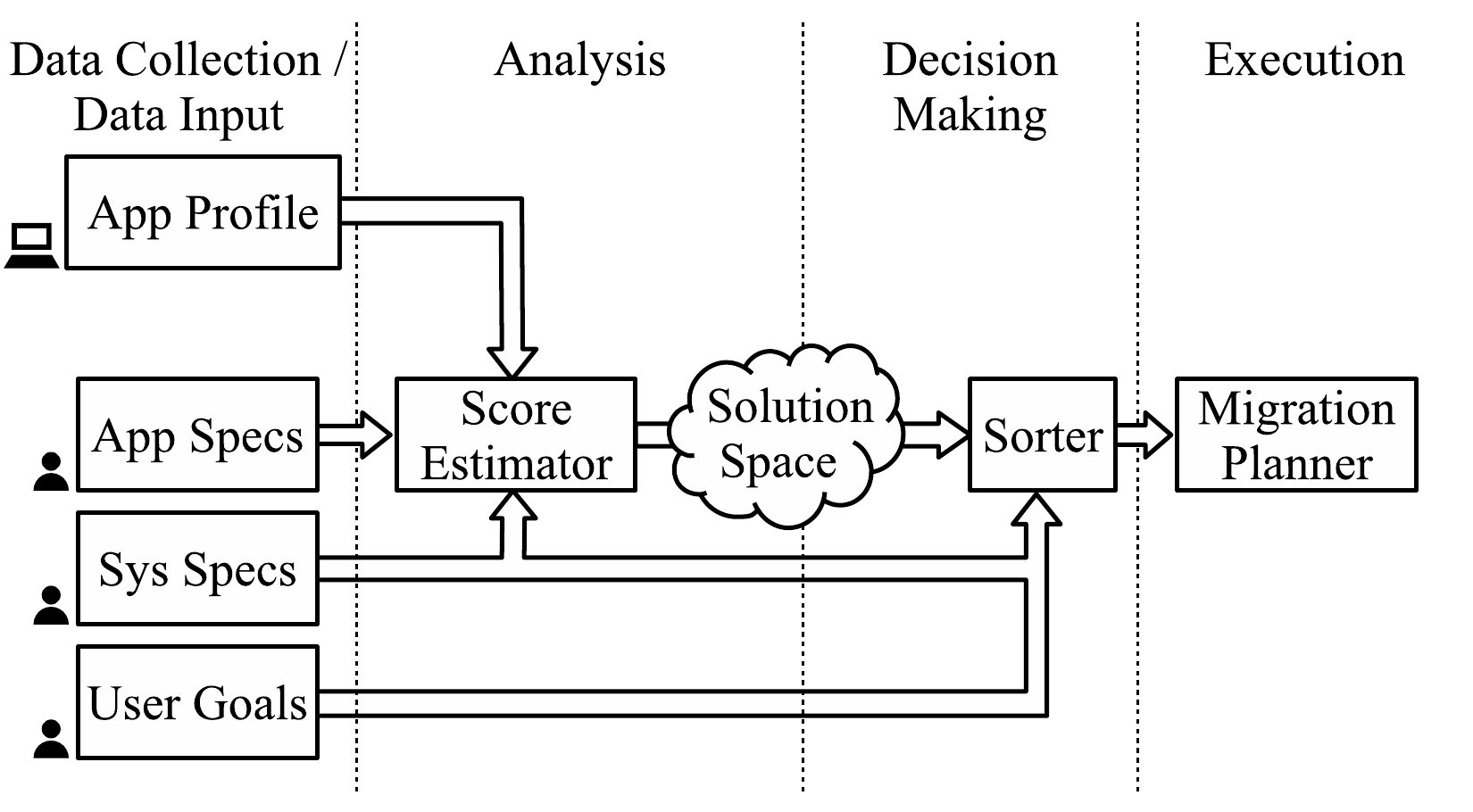}
			\caption{General storage tiering phases}
			\label{fig:tiering}
		\end{figure}
	
		Figure~\ref{fig:tiering} illustrates a general storage tiering mechanism which consists of four phases. In the data collection phase, the system gathers required information for decision making. 
		
		The application profile of IO pattern can be obtained either online or offline.
		An online profiling module may collect IO information while the application is running at the cost of potential performance overhead. This mechanism is useful when there is a user involved, such as personal computers, or virtual environment cloud systems. An offline profiling module, on the other hand, obtains the application IO profile before it is running. This kind of profiling mechanism is suitable for cluster analytical applications in which no random parameter interferes with the IO path except the running applications which are predictable. Some other information such as application/system specifications can also be fed into the tiering algorithm by the user or the machine all at once.
		In the analysis phase, the system evaluates several possible plans or models and generates a list of recommendations in the form of a solutions space. Some tiering algorithms may skip this phase by directly finding the answer with some analysis. The solution space consists of several estimations under different circumstances evaluated by a cost function or a performance model (e.g., running a particular application or the whole system under a particular distribution of data among the tiers). Each solution comes with a cost estimation which will be later used for decision making in the next phase.
		In this phase, a sorting algorithm might suffice for deciding which migration plan is worth taking. According to the goals, the scores of each plan, and their costs, a tiering algorithm determine whether or not migrating a chunk of data in which direction.
		
		\subsection{SSD as a Performance Tier}
		
		A comprehensive study on available storage types is provided in \cite{kim2014evaluating}. It compares the Micron all-PCM SSD prototype with eMLC flash SSD regarding performance and evaluates it as a promising option for tiered storage systems. Using a simulation methodology with estimated/obtained performance characteristics of each device, it tests every possible combination of PCM SSD, eMLC SSD, and HDD. Although nowadays we have Optane SSD available in the market from Intel and Micron, and we know that it offers much better performance than the out-of-date all-PCM SSD prototype, this paper assumes that the write operation of PCM SSD is 3.5x slower than that of eMLS SSD. With this assumption, and having a very simple IOPS based dynamic tiering algorithm, they show the benefits of using PCM SSD in a multi-tiered storage system in a variety of real-world workloads as an enterprise solution.
		
		Online OS-level Data Tiering (OODT)~\cite{salkhordeh2015operating} efficiently distributes data among multiple tiers based on the access frequency, data type (metadata or user data), access pattern (random or sequential), and the read/write operation frequency. Using a weighted priority function, OODT sorts data chunks for each tier based on their degree of randomness, read ratio, and request type. OODT can interpret fixed size requests (4KB). If the request is larger than that, it will be broken into several small sub-requests and treat with them independently in a module called the dispatcher. By using a mapping table, all data chunks can be tracked down to the tier number and the physical block index. To enable online migration, it obtains the statistics of the blocks and keeps it in an access table which gets up-to-date by the dispatcher. The most important part of OODT, and every other tiering schemes is the priority computation (may be referred as scoring, sorting, or matching in other techniques) which determines the matched tier for each data. Using a simple weighted linear formula with four inputs of $P_{access}$, $P_{random}$, $P_{read}$, and $P_{metadata}$, OODT calculates the priority for potential migrations.
		
		Cloud Analytics Storage Tiering (CAST)~\cite{cheng2015cast}, as it sounds, is a storage tiering solution for data analytics applications in the cloud. With an offline workload profiling, CAST makes job performance prediction models for each tenant on different cloud storage services. Then, it combines the obtained prediction models with workload specifications and its goals to perform a cost-efficient data placement and storage provisioning plan. They model the data placement and storage provisioning problem into a non-linear optimization problem in which they maximize the tenant utilization in terms of the completion time and the costs. An enhanced version of CAST is also proposed in \cite{cheng2015cast} which is called {CAST++} and adds data reuse patterns and workflow awareness to CAST.
		
		Based on a measured IOPS in a virtualization environment, AutoTiering~\cite{yang2017autotiering} dynamically migrates virtual disk files (VMDK) from one tier to another in an all-flash storage system. It uses a sampling mechanism to estimate the IOPS of running a VM on other tiers. Based on this measurement and the costs, it sorts all possible movements by their scores. For each VMDK, a daemon on the hypervisor collects the IO related statistics including the IOPS results of latency injection test to resemble a slower tier at the end of every sampling period. For simulating faster tiers, AutoTiering takes benefits of a linear regression model. If the IOPS does not change by slowing down the IO process, and there is a VM in the queue waiting for the performance tier, a demotion will take the VMDK to a capacity tier and let the other VMDK take over the performance tier by promoting it.
		
		\begin{figure*}[t]
			\centering
			\includegraphics[width=7in]{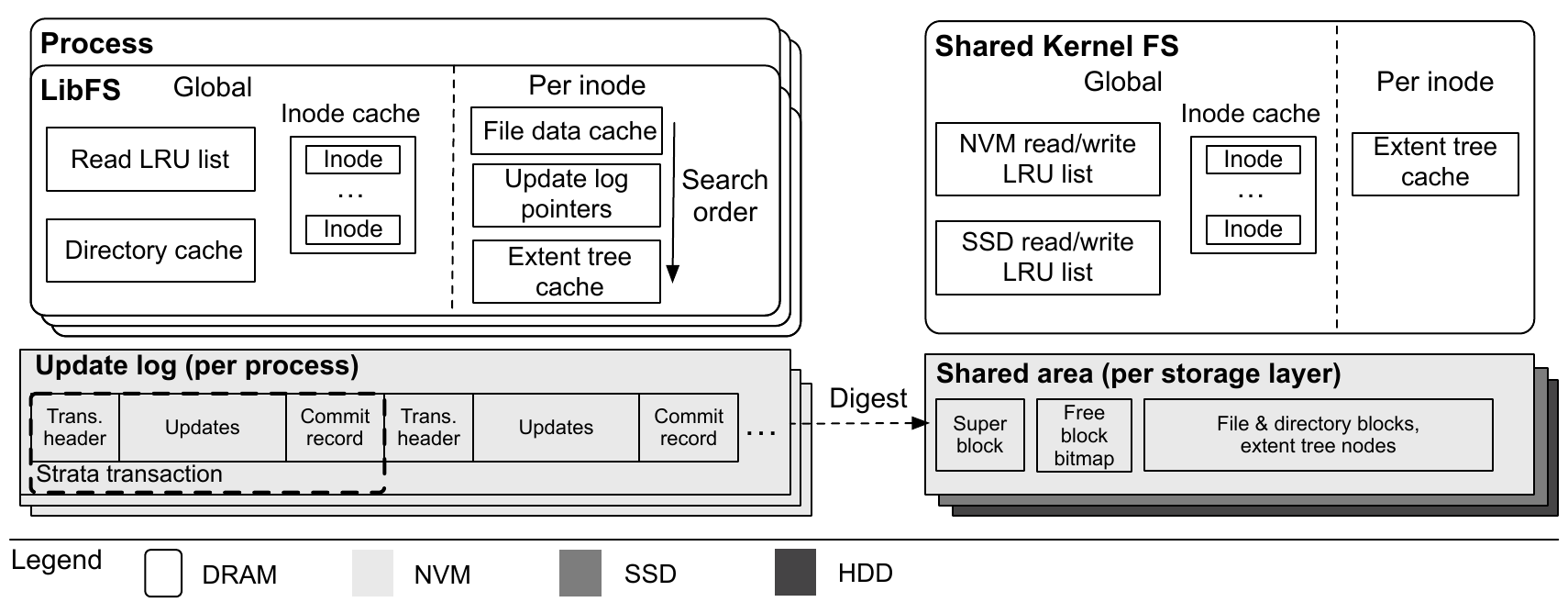}
			\caption{Strata design~\cite{kwon2017strata}. LibFS directs writes to the \textit{update log} and serve the reads from the shared area. File data cache is a read-only cache, containing data from SSD or HDD.}
			\label{fig:strata}
		\end{figure*}
		
		\subsection{NVM as a Performance Tier}
		
		NVMFS~\cite{qiu2013nvmfs} is a hybrid file system that improves the random write operations in NAND-flash SSD by exploiting the byte-addressability of an auxiliary NVM device. The key feature of this file system is that it redirects small random IOs on NVM which include metadata and hot file data blocks. This scheme helps to reduce the write traffic to SSD, hence improves SSD's durability. The technique is to transform random writes at the file system level to a sequential write at the SSD level. It groups data with same update likelihood and submits a single large SSD write request.
		
		NVMFS comprises 2 LRU lists: dirty and clean. The dirty LRU list absorbs updates in the NVRAM. When a page is written back to the SSD device, it moves from dirty list to the clean list. NVMFS dynamically adjust dirty and clean LRU lists. Once the NVRAM usage reaches 80\%, a background thread starts flushing data from the dirty list and move them to the clean LRU list until it goes down to 50\%. NVRAM has a non-overwrite on SSD policy: periodical cleanup internal fragmentation that integrates multiple partial extents into one and recycles the free space. 
		
		The authors explain file system consistency through 5 steps. 1: Check if the NVRAM usage is over 80\%; 2: if so, group random small IOs from the dirty LRU list into large (512K) extents; 3: then, sequentially write the extent to SSD (better block erase at FTL); 4: insert the flushed pages into the clean LRU list; and finally 5: update metadata by recording the new data position within \verb+page_info+ structure. Therefore, when a crash happens at any point, it can be recovered.
		
		To prevent segment cleaning inconsistency, NVMFS exploits transactions during defragmentation, similar to the transactions in log-structured file systems. After choosing a candidate extent, it migrates the valid blocks of that to NVRAM, and then updates the corresponding \textit{inodes}. When the inodes are updated, then it releases the space in SSD. Data will always be consistent even when a crash happens in the middle of the process.
		
		Strata~\cite{kwon2017strata} is a multi-tiered file system which exploits NVM as the performance tier, and SSD/HDD as the capacity tiers.
		It consists of two parts: KernFS and LibFS.
		To fire up Strata, applications are required to be recompiled with LibFS which re-implements standard POSIX interface.
		On the kernel side, KernFS should be running to grant the application access to the shared storage area which is a combination of NVM, SSD, and HDD. It uses the byte-addressability of the NVM to coalesce logs and migrate them to lower tiers to minimize write amplification. File data can only be allocated in NVM in Strata, and they can be migrated only from a faster tier to a slower one. The profiling granularity of Strata is a page, which increases the bookkeeping overhead and wastes the locality information of file accesses.
		
		Strata attains fast write operation by separating and delegating the tasks of logging and digesting to the user space and the kernel space, respectively. The KernFS grants LibFS direct access to the NVM for its own private \textit{update log} and the the \textit{shared area} for read-only operations, as shown in figure~\ref{fig:strata}. KernFS perform the digest operation in parallel via multiple threads. One benefit of this operation is that despite the randomness and small size of the initial writes to the update log, they can be coalesced and written sequentially to the shared area which helps to minimize fragmentation and metadata overhead. This also helps efficient flash erasure and shingled write operations. 
		
		For crash consistency, LibFS works with a durable unit of \textit{Strata transaction} which provides ACID semantics to the applications update log. To implement this, Strata wrap every POSIX system call in one or multiple Strata transactions. Figure~\ref{fig:strata} represents the Strata design and the LibFS and KernFS components.

		\subsection{NVM as a Metadata Tier}
		
		In a journaling file system, like Ext4, the metadata updates are usually very small (e.g. \textit{inode} size of 256B). Although modifying an \textit{inode} requires small write operations, due to block size operations of the storage devices, a whole \textit{inode} block (e.g., 4K) would be replaced. In recent years, Non-Volatile Memories have attracted a lot of attention due to their feature of connecting via the memory bus. This feature means that the CPU may issue byte-level (cache line size) persistent updates.
		
		File System Metadata Accelerator (FSMAC)~\cite{chen2013fsmac} decouples data and metadata I/O path and use NVM to store file system metadata due to its small access size and popularity. Metadata is permanently stored in NVM and by default, never flushed back to the disk periodically. All updates to the metadata are in-place updates. Not only after a power failure in the middle of a metadata update operation, metadata would be corrupt, but also the authors argue that because of the performance gap between NVM and a block device, the data update is behind metadata update which becomes persistent in NVM once updated. Since the byte-size versioning is very complex and tedious to implement, and block-size versioning imposes write amplification and NVM space waste, FSMAC uses fine-grained versioning (\textit{inode}-size, i.e., 128bytes) that can maintain consistency at reasonable implementation and space costs. 
		
		To address the write ordering issue of data and metadata without destroying the performance gained due to the byte-addressability of NVM, FSMAC uses a light-weight combination of fine-grained versioning and transaction. An original version of metadata is created before updating it to recover from a crash securely. It will be deleted only after the successful completion of the updating transaction. After that, the whole file system will be consistent.
		
		Using this opportunity, C. Chen et al. proposed fine-grained metadata journaling on NVM~\cite{chen2016fine}. Although it is not directly related to tiering nor caching solution, using NVM to keep a part of storage data is a kind of classification problem which is fundamental in tiering approaches.
		
		\begin{figure}[h]
			\centering
			\includegraphics[width=3.5in]{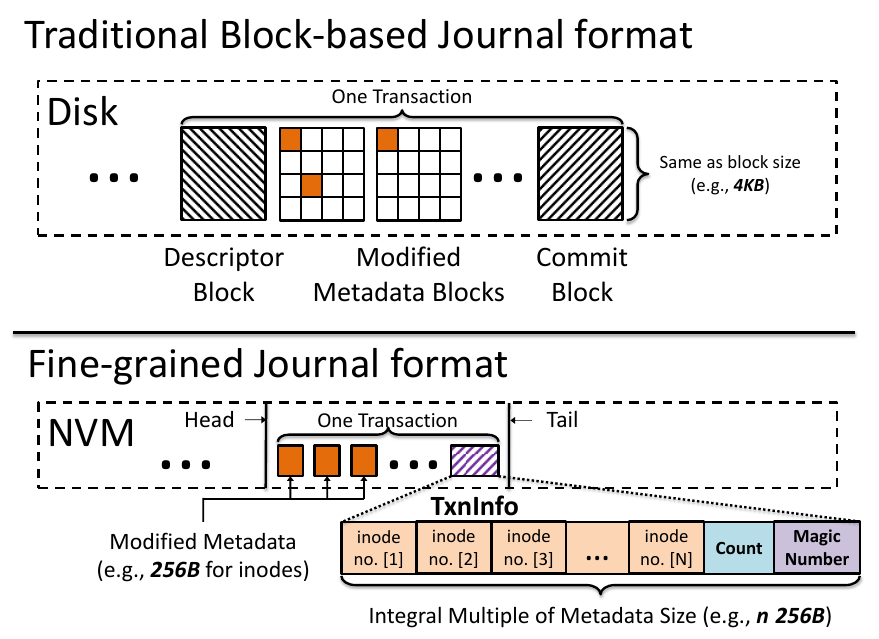}
			\caption{Fine-grained Metadata Journal Format on NVM~\cite{chen2016fine}}
			\label{fig:fine-grnd}
		\end{figure}
		
		In contrast to conventional journaling file systems in which the modified \textit{inode} blocks in the page buffer in DRAM are persisted to the disk in form of transactions, in NVM-base fine-grained journaling file system~\cite{chen2016fine}, only modified \textit{inodes} are linked together and persisted in the NVM (Figure~\ref{fig:fine-grnd}). Using cache flush instruction and memory fence, it provides the consistency of ordered writes. Instead of using large Descriptor and Commit (or Revoke) blocks which are 8K in total, a new data structure, \verb+TxnInfo+, is introduced which contains the number of modified \textit{inodes} in the list (\verb+Count+), and a \textit{Magic Number} for identifying \verb+TxnInfo+ during the recovery time.
		
		The journal area in NVM is a ring buffer with a head and a tail pointer. Writing to it is composed of three steps: 1) \verb+memcpy+ modified \textit{inodes} from DRAM to NVM; 2) flush the corresponding cache lines and issue a memory barrier; and 3) atomically update the tail pointer in the journal area in NVM using the atomic 8-byte write, flush its cache line, and issue a memory barrier.

		\begin{figure*}[t]
			\subfloat[The file structure of Ziggurat and basic migration\label{fig:mig_intro}]{
				\includegraphics[height=5.4cm]{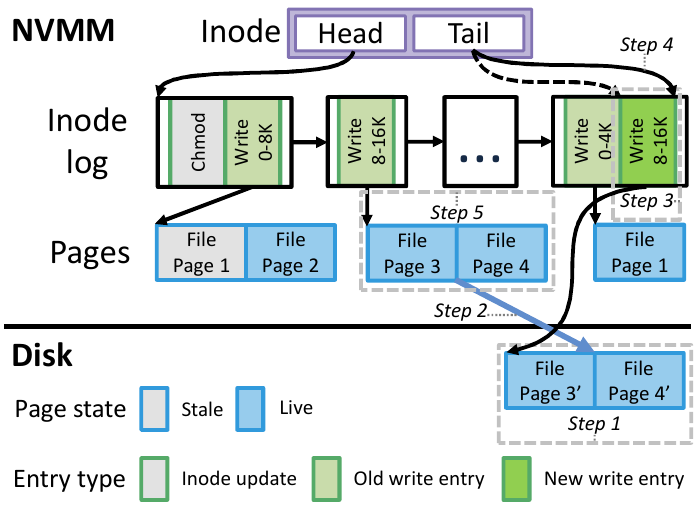}
			}
			\subfloat[Group migration\label{fig:mig_range}]{
				\includegraphics[height=5.4cm]{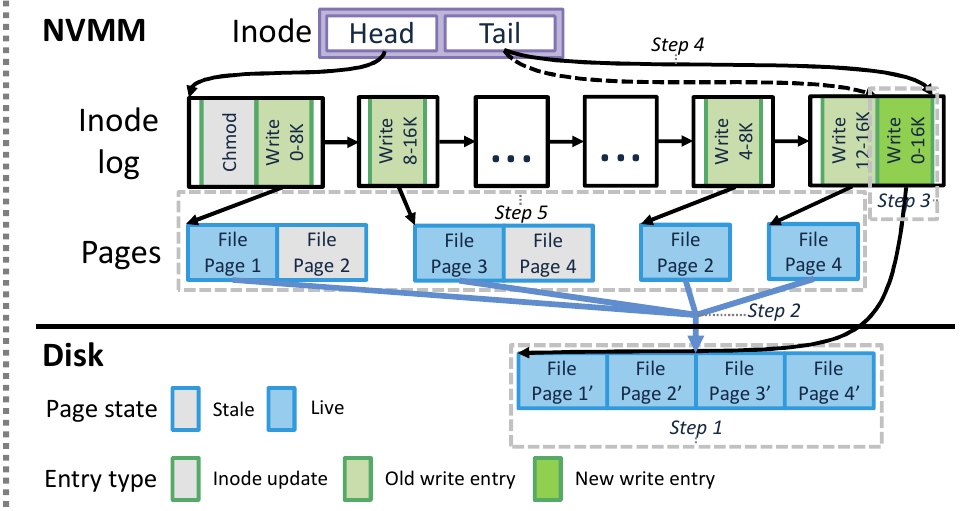}
			}
			\caption{\textbf{Migration mechanism of Ziggurat}~\cite{shengan2019ziggurat}.
				Ziggurat migrates file data between tiers using its basic migration and group migration mechanisms.
				The blue arrows indicate data movement, while the black ones indicate pointers.}
			\label{fig:mig}
		\end{figure*}
		
		In traditional journaling file systems, committing the \textit{Running Transaction}, which is a link list of modified \textit{inode} blocks, is triggered by either a predefined timer or a predefined number of modified \textit{inode} blocks. In the fine-grained journaling, when a predefined timer is up, similar to traditional file systems, the committing process starts. The unset of this process is also controlled by the number of modified \textit{inodes}, because \verb+TxnInfo+ can hold the information of a limited number of modified \textit{inodes}. The committing process begins with relinking all modified \textit{inodes} from the \textit{Running Transaction} to the \textit{Committing Transaction} so that the \textit{Running Transaction} can accept new modified \textit{inodes}. Then, all modified \textit{inodes} are \verb+memcpy+ed to NVM starting from tail, and then the \verb+TxnInfo+ is calculated afterwards. The corresponding cache lines, thereafter, are flushed, and a memory fence is issued. Finally, the tail pointer will atomically get updated, confirming that the transaction is committed. Notice that, data is consistent during this process, even with a crash happening in the middle, because the tail is controlling the visibility of data. Comparing to traditional journaling, this method reduces transaction writes by up to 99\%.
		
		To prevent too long journals which deteriorates the performance, file systems usually use checkpointing periodically. The fine-grained journaling triggers checkpointing once in every 10 minutes or upon 50\% utilization of the NVM. Like traditional journaling file systems, it takes over the modified \textit{inode} block list and write the blocks one after another. Then, it discards the journal in NVM by making head and tail pointers equal, which guarantees the recoverability, because when a crash happens, we still have the modified \textit{inodes} in NVM in the recovery.
		
		The recovery process in the fine-grained journaling starts from the tail in NVM, backward. It retrieves the corresponding \textit{inode} blocks to DRAM. The obsolete \textit{inode} blocks get up-to-date by applying modified \textit{inodes} inside the block. After that all \textit{inode} blocks are updated in DRAM, it flushes them back to the disk. Finally, make the head and tail pointers identical atomically. The consistency is guaranteed similar to the checkpointing process.

		
		\subsubsection{Our Multi-Tiered File System: Ziggurat\protect\footnote{Parts of this section is taken from the original paper accepted in FAST'19~\cite{shengan2019ziggurat}}}
		\label{sec:zig}
		
		Ziggurat~\cite{shengan2019ziggurat} is a multi-tiered NVM-based file system a tiered file system that spans NVNM and disks, and it was developed in our research group. The paper is published in the proceedings of the 17th USENIX Conference on File and Storage Technologies (FAST '19). It is based on our well-known NVM-base file system, NOVA~\cite{xu2016nova}. Ziggurat exploits the benefits of NVM through intelligent data placement during file writes and data migration. 
		Ziggurat includes two placement predictors that analyze the file write sequences and predict whether the incoming writes are both large and stable and whether updates to the file are likely to be synchronous.
		Then, it steers the incoming writes to the most suitable tier based on the prediction: writes to synchronously-updated files go to the NVM tier to minimize the synchronization overhead.
		Small, random writes also go to the NVM tier to entirely avoid random writes to disk. 
		The remaining large sequential writes to asynchronously-updated files go to disk. Ziggurat seeks five principal design goals which are as follows.
		
		\textbf{Send writes to the most suitable tier}. Although NVM is the fastest tier in Ziggurat, file writes should not always go to NVM. 
		NVM is best-suited for small updates (since small writes to disk are slow) and synchronous writes (since NVM has higher bandwidth and lower latency). 
		However, for larger asynchronous writes, targeting disk is faster, since Ziggurat can buffer the data in DRAM more quickly than it can write to NVM, and the write to disk can occur in the background. 
		Ziggurat uses its synchronicity predictor to analyze the sequence of writes to each file and predict whether future accesses are likely to be synchronous (i.e., whether the application will call \texttt{fsync} shortly).
		
		\textbf{Only migrate cold data in cold files}. During migration, Ziggurat targets the cold portions of cold files. 
		Hot files and hot data in unevenly-accessed files remain in the faster tier.
		When the usage of the fast tier is above a threshold, Ziggurat selects files with the earliest average modification time to migrate. 
		Within each file, Ziggurat migrates blocks that are older than average. Unless the whole file is cold (i.e., its modification time is not recent), in which case we migrate the entire file.
		
		\textbf{High NVM space utilization}. Ziggurat fully utilizes NVM space to improve performance. Ziggurat uses NVM to absorb synchronous writes. Ziggurat uses a dynamic migration threshold for NVM based on the read-write pattern of applications, so it makes the most of NVM to handle file reads and writes efficiently. We also implement reverse migration to migrate data from disk to NVM when running read-dominated workloads.
		
		\textbf{Migrate file data in groups}. To maximize the write bandwidth of disks, Ziggurat performs migration to disks as sequentially as possible. The placement policy ensures that most small, random writes go to NVM. However, migrating these small write entries to disks directly will suffer from the poor random access performance of drives. To make migration efficient, Ziggurat coalesces adjacent file data into large chunks for movement to exploit sequential disk bandwidth. 
		
		\textbf{High scalability}. Ziggurat extends NOVA’s per-CPU storage space allocators to include all the storage tiers. It also uses per-cpu migration and page cache write-back threads to improve scalability.
		
		Figure~\ref{fig:mig_intro} shows the basic procedures of how Ziggurat migrates a write entry from NVM to disk. The first step is to allocate continuous space on disk to hold the migrated data. Ziggurat copies the data from NVM to disk. Then, it appends a new write entry to the inode log with the new location of the migrated data blocks. After that, it updates the log tail in NVM and the radix tree in DRAM. Finally, Ziggurat frees the old blocks of NVM.
		
		Figure~\ref{fig:mig_range} exhibits the steps of group migration which avoids fine-grain migration to improve efficiency and maximize sequential bandwidth to disks. They are similar to migrating a write entry. In step 1, it allocates large chunks of data blocks in the lower tier. In step 2, it copies multiple pages to the lower tier with a single sequential write. After that, it appends the log entry, and update the \textit{inode} log tail, which commits the group migration. The old pages and logs are freed afterward. Ideally, the group migration size (the granularity of group migration) should be set close to the future I/O size, so that applications can fetch file data with one sequential read from disk. Also, it should not exceed the CPU cache size to maximize the performance of loading the write entries from disks.
		
		In a nutshell, Ziggurat bridges the gap between disk-based storage and NVM-based storage and provides high performance and large capacity to applications.
	
	\section{Conclusion}
	\label{sec:conc}
		The diversity in storage technologies and their different characteristics make each of them individually suitable for a set of storage needs. In the software side, the ever expanding cloud of digital information requires large scale enterprise data servers with high-performance storage systems. While old well-designed storage technologies like HDDs provide large space and high density at a relatively low costs, new technologies such as SSD and NVM offer super fast and reliable IO workflow at a much higher costs. The general desire is to have the high-performance of the new technologies with high storage capacities and low costs. Despite the fact that the speed of processor's development is much higher than the storage technology development, software solutions such as caching and tiering attract the expert's attention to overcome the aforementioned limitation. In this survey, we extensively investigated several caching and tiering solutions for high-performance storage systems. 
		We observed that although there are several caching and tiering proposals which use SSD as the performance tier, the young technology of NVM did not receive enough attention to be used in such systems. It is not unexpected since this technology has been developed recently and the first products of this type is shipped to the market just a few month before this publication. By the way, we also looked into some recent scientific papers on using NVM as a performance tier, and we also introduced Ziggurat, a multi-tiering file system using NVM as a performance tier to cover the long latencies of SSDs and HDDs.

	{	
		\normalsize 
		\bibliographystyle{acm}
		\bibliography{references}
	}

	
\end{document}